# Temperature-assisted Piezoresponse Force Microscopy: Probing Local Temperature-Induced Phase Transitions in Ferroics

Anna N. Morozovska[1], Eugene A. Eliseev[2], Kyle Kelley[3], Sergei V. Kalinin[3,*],

[1] Institute of Physics, National Academy of Sciences of Ukraine, 46, pr. Nauky, 03028 Kyiv, Ukraine

[2] Institute for Problems of Materials Science, National Academy of Sciences of Ukraine, Krjijanovskogo 3, 03142 Kyiv, Ukraine

[3] Center for Nanophase Materials Sciences, Oak Ridge National Laboratory, Oak Ridge, TN 37922

Combination of local heating and biasing at the tip-surface junction in temperature-assisted piezoresponse force microscopy (tPFM) opens the pathway for probing local temperature induced phase transitions in ferroics, exploring the temperature dependence of polarization dynamics in ferroelectrics, and potentially discovering coupled phenomena driven by strong temperature- and electric field gradients. Here, we analyze the signal formation mechanism in tPFM and explore the interplay between thermal- and bias-induced switching in model ferroelectric materials. We further explore the contributions of the flexoelectric and thermopolarization effects to the local electromechanical response, and demonstrate that the latter can be significant for "soft" ferroelectrics. These results establish the framework for quantitative interpretation of tPFM observations, predict the emergence the non-trivial switching and relaxation phenomena driven by non-local thermal gradient-induced polarization switching, and open a pathway for exploring the physics of thermopolarization effects in various non-centrosymmetric and centrosymmetric materials.

---

* Corresponding author: sergei2@ornl.gov

## I. INTRODUCTION

For over a century, ferroelectric and polar materials have remained one of the central research areas in condensed matter physics and materials science [1]. Initial interest in these materials was driven by the applications of ferroelectric single crystals and poled ceramics in sonars, sensors and actuators, and electrooptical imaging [2, 3, 4]. At the same time, the progress in thin film deposition technologies in the 80's and 90's spurred interest in applications of ferroelectrics in information storage and processing devices including non-volatile memories [5, 6], FETs with ferroelectric gates [7], and tunneling junctions [8, 9]. In parallel, much attention has been focused on the fundamental physics of ferroelectric materials driven by the interplay between local ferroelectric instabilities and non-local depolarization phenomena [10, 11].

Note that until 2010, the ferroelectric research was dominated by classical perovskite oxides [12], with small niche studies of materials such as water soluble triglycine sulphate and Rochelle salt, preponderantly for technique development. However, the situation has changed drastically over last decade due to the appearance of binary ferroelectrics such as $HfO_2$ [13, 14], $Zn_xMg_{1-x}O$ [15], $Al_xB_{1-x}N$ [16, 17]; improper ferroelectrics such as $ErMnO_3$ [18, 19], boracites [20, 21, 22], hybrid perovskites [23]; low-dimensional layered van der Waals ferroelectrics such as $CuInP_2S_6$ [24, 25, 26] and related materials, $Sn_2P_2S_6$ [27] and $In_2Se_3$; [28] and particular twisted 2D $MX_2$ structures [29, 30]. At the same time, novel phenomena were discovered in classical and emergent ferroelectric materials, including pressure-induced switching [31], strong coupling between the electrochemical and ionic phenomena at surfaces and interfaces [32, 33, 34], chemical switching [35, 36], and emergence of ferroionic [37, 38] and antiferroionic states [39], and others. Furthermore, some classes of materials such as 1D ferroelectric SbSI remain virtually unexplored by the ferroelectric community [40, 41].

These considerations necessitate exploring the fundamental phenomena in ferroelectrics, including the domain structure and polarization switching, interaction between topological and structural defects, and topological defect dynamics. Similarly, of interest is the coupling between ferroelectricity and other functionalities, such as elasticity, conductivity, etc. Common for virtually all material classes is that these phenomena are highly local, necessitating the development of methods for probing these phenomena on the nanometer scales of domain walls, structural and topological defects, or Moire lattice in twisted bilayers.

Over the last 25 years, the piezoresponse force microscopy (**PFM**) has become the technique of choice for the nanoscale probing and modification of ferroelectric materials [42, 43, 44, 45]. PFM imaging is based on the detection of minute displacements in polar materials induced by the application of bias to the PFM tip. The electromechanics of the tip-surface junction in PFM has been analyzed by multiple authors in the uniform field [46] and decoupled [47, 48, 49] approximations. Similarly, full analytical solutions for the coupled piezoelectric indentation problem have been derived for certain materials symmetries [50, 51]. It has been demonstrated that under general conditions, the PFM signal is independent of the poorly controlled tip-surface contact radius, rendering this technique intrinsically quantitative [52]. Similarly, the direct and converse electromechanical effects at the tip-surface junction were shown to be equivalent [53], opening the pathway for probing fundamental physics of polar materials.

The further development of PFM has led to the emergence of a broad gamut of voltage- and time spectroscopies [54]. In these techniques, the tip is fixed at a certain position and its bias is varied as a function of time [55, 56]. In ferroelectrics, the application of bias to the probe can induce nucleation and growth of the ferroelectric domains of opposite polarity. Hence, the bias evolution of the signal, i.e., local hysteresis loops, provides insight into the domain nucleation and growth process [57, 58, 59]. For materials with more complex functionalities, such as ferroelectric relaxors or electrochemical systems, the mechanisms behind the bias- and time dependence of electromechanical response are more complex [60, 61]. However, in all cases when the bias-induced changes are reversible, the measurements can be performed over the spatial grid of points, providing insight into the spatial variability of the bias-induced phase transformations and reactions [62]. Note that implicitly, many of these advances are enabled by the fact that the signal in PFM is independent of contact area, and hence measured responses are dominated by the intrinsic materials variability. These techniques further necessitate the development of models for signal evolution with bias or time [10, 63, 64]. However, once accomplished this allows the transition from largely qualitative observations and spectroscopies common for SPM to quantitative studies of bias-induced phase transitions in nanoscale volumes [59, 65, 66, 67].

Furthermore, understanding ferroelectric phenomena necessitates probing temperature-induced and temperature-dependent processes. To gain insight into temperature-induced phenomena in ferroelectrics, a number of groups have explored the evolution of the PFM signal and complementary Kelvin Probe Force Microscopy signals under global heating across the phase transitions [68, 69, 70, 71]. A number of interesting phenomena including the domain branching and domain memory effects in PFM, temperature induced potential inversion [72] and potential retention, and relaxation above the Curie temperature [73] were reported and attributed to the external screening of polarization charges. In this process, the polarity of the observed ferroelectric domains can be opposite to the intrinsic polarization charge sign. Interestingly, a number of early observations, such as the PFM signals temperature dependence that scales with polarization (rather then diverge as expected for piezoelectric constant) [74], still remain unexplored.

However, the extant implementations and analyses of PFM are limited to uniform temperature observations and, in few cases, to observation of domain structures under macroscopic thermal gradients [75]. This severely limits the range of ferroelectric phenomena that can be explored. Much like how capacitor-based PFM measurements activate all defects [76, 77, 78, 79], global heating leads to global changes in domain structures. For example, if one of the defects has a lower transition temperature, it introduces a phase transition in a macroscopic volume precluding exploration of other (weaker) defect centers. This general limitation precluded systematic studies of temperature-induced phase transitions or coupled thermal and bias-induced phenomena. Secondly, much like how strain gradients give rise to a broad range of flexoelectricity-driven phenomena, it can be expected that sharp local temperature gradients will reveal flexo- and thermopolarization effects.

Recently, advances in SPM instrumentation have allowed the combined imaging and spectroscopy modes, when both the temperature and the bias of the probe can be varied [80, 81]. In this fashion, the local biasing and local heating of the microscopic volumes of the material at the tip-surface junction can be affected simultaneously. Here we analyze the mechanisms of ferroic interaction with a heated PFM tip, explore the evolution of bias- and temperature induced polarization distributions, and derive the temperature-dependent responses. We calculate the solution of a thermo-elastic-electric probing problem fully coupled with Landau-Ginzburg-Devonshire (**LGD**) description of the ferroelectric polar properties, and analyze the temperature-induced and voltage-induced polarization redistribution and local electromechanical response occurring under the heated PFM tip.

The manuscript is structured as following. **Section II** contains the formulation of the local thermo-elastic-electric probing problem with boundary conditions and material parameters used in calculations. **Section III** analyzes the temperature-induced polarization redistribution, elastic strains and surface displacement of a ferroelectric layer at zero voltage applied to the heated PFM tip. The changes of the ferroelectric polarization and local electromechanical response induced by the biased and heated PFM tip are considered in **Section IV**. Lastly, **Section V** is a brief summary. Calculation details and auxiliary figures are listed in **Suppl. Mat**. [82].

## II. THEORETICAL DESCRIPTION
### A. Coupled thermo-elastic-electric problem for a ferroelectric layer

Here we consider the case of a scanning probe microscopy tip in contact with the ferroelectric surface, common for contact mode scanning probe microscopies. The geometry of calculations is conventional for PFM probing and is shown in **Fig. 1(a)-(c)**. The free energy dependences on the polarization $P$ and temperature $T$ calculated for the "soft" and "hard" uniaxial ferroelectrics, $Sn_2P_2S_6$ and $LiNbO_3$, are shown in **Fig. 1(d)** and **1(e)**, respectively. It is evident how the change in temperature changes the spontaneous polarization wells. Applied voltage makes these wells non-equivalent up to the disappearance of the shallow well at coercive voltage.

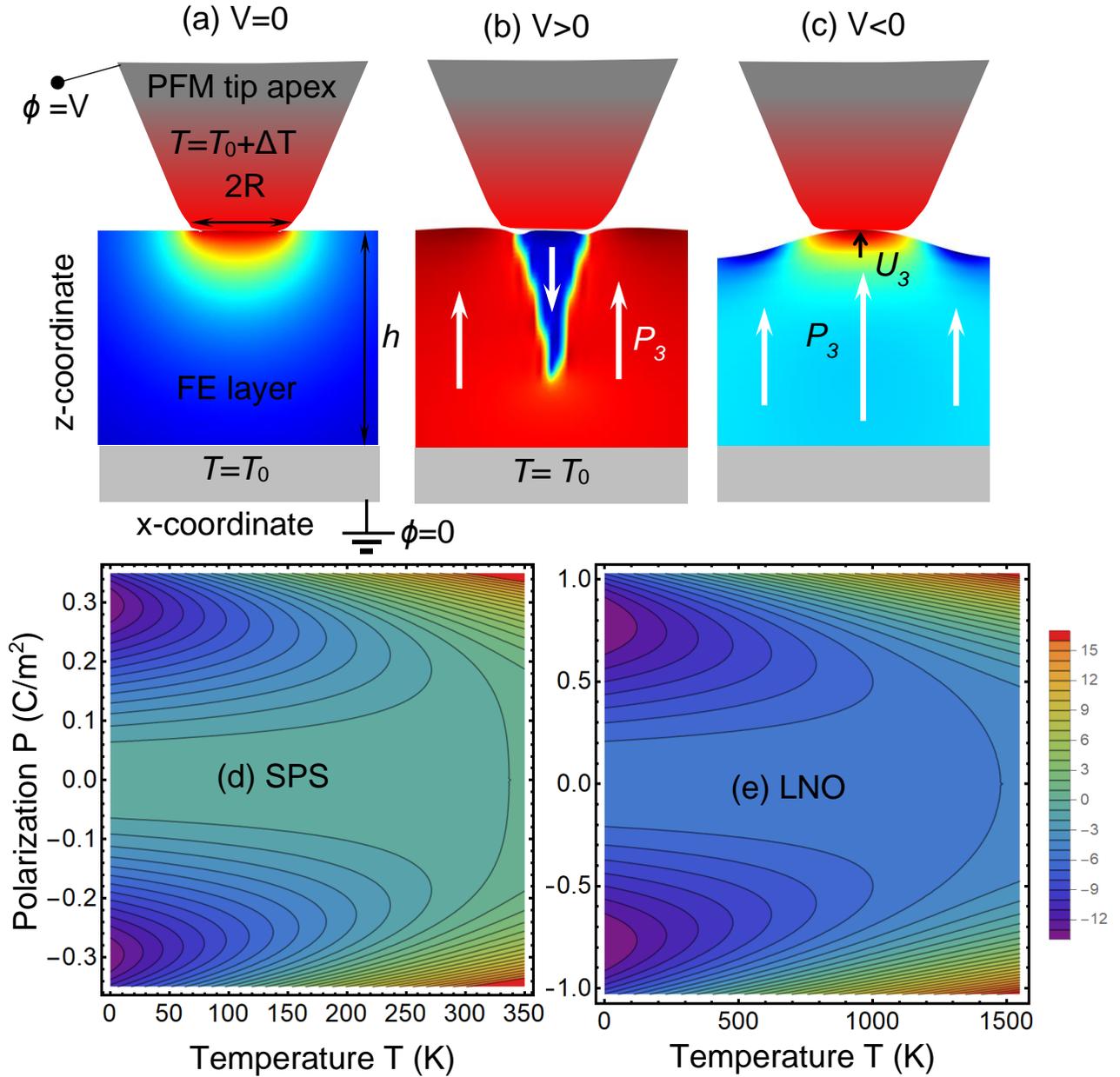

**FIG. 1.** (a) The temperature distribution in a FE layer of thickness $h$ induced by a PFM tip heated up to the temperature $T_0 + \Delta T$. The temperature of remote bottom electrode is $T_0$. Polarization distribution in the FE layer under the PFM tip biased with either positive (b) or negative (c) voltage, respectively. The free energy dependence on the polarization $P$ and temperature $T$ calculated for the "soft" and "hard" uniaxial ferroelectrics, $Sn_2P_2S_6$, (d) and $LiNbO_3$ (e), respectively.

Here, we assume that the temperature distribution obeys a standard heat equation. The heated tip apex, the FE layer and the ambient are characterized by their own thermal conductivity equation for the temperature variation $T_m(\vec{r}, t)$ inside each region "$m$":

$$\frac{\partial}{\partial t} T_m = \kappa_m^T \left(\frac{\partial^2}{\partial x^2} + \frac{\partial^2}{\partial y^2} + \frac{\partial^2}{\partial z^2}\right) T_m. \qquad (1)$$

The FE layer occupies the region $0 \leq z \leq h$. The coefficients $\kappa_m = k_m^T/c_m$, where $c_m$ is the heat capacity and $k_m^T$ is the thermal conductivity of the medium "$m$". The relation between the heat flux and the temperature variation is given by the conventional expression, $\vec{j}_m = -k_m^T \frac{\partial \vartheta_m}{\partial \vec{n}}\Big|_{S_m}$. Thermal boundary conditions to Eq.(1) at their physical boundaries $S_m$ are the continuity of heat fluxes and the equality of the media temperatures (see **Appendix A** for details).

Due to the very high heat conductivity of the metallic tip, moderate conductivity of a FE layer and very low ambient conductivity, one can neglect the heat flux between the FE and the ambient air or vacuum. Here we assume that the tip is heated by $\Delta T$, is in contact with a FE layer, which temperature is $T_0$ very far from the surface $z = 0$, e.g., at the remote bottom electrode $z \geq h$. The tip apex is modeled as a perfectly heat-conducting disk with effective radius $R$ being in a tide electric and thermal contact with the FE surface $z = 0$. This model corresponds to a well-known disk-plane model of the PFM tip [83, 84], where $R$ can be associated with the contact radius of the tip apex. Note that here we assume the temperature-induced changes in polarization affect temperature field and dynamics only weakly, i.e., adopt decoupled approximation for the thermal field.

In order to find the spatial distribution of the acting electric field $E_i$ and out-of-plane ferroelectric polarization component $P_3$ inside a uniaxial FE, one should solve a coupled problem consisting of Poisson equation for electric potential $\varphi$ and LGD-type equation for $P_3$:

$$\left(\frac{\partial^2}{\partial x^2} + \frac{\partial^2}{\partial y^2} + \frac{\partial^2}{\partial z^2}\right) \phi^{(in)} = \frac{1}{\varepsilon_0 \varepsilon_b} \frac{\partial P_3}{\partial z}, \qquad (2a)$$

$$[\alpha_T(T(\vec{r}) - T_C) - Q_{ij33}\sigma_{ij}]P_3 + \beta P_3^3 + \gamma P_3^5 - g_{11}\frac{\partial^2 P_3}{\partial z^2} - g_{44}\left(\frac{\partial^2 P_3}{\partial x^2} + \frac{\partial^2 P_3}{\partial y^2}\right) = \mu \frac{dT}{dz} + E_3 - F_{ijk3}\frac{\partial \sigma_{ij}}{\partial x_k}. \qquad (2b)$$

Here $\varepsilon_b$ is a background permittivity [85], $T(\vec{r})$ obeys Eq.(1), $T_C$ is a bulk Curie temperature, $Q_{ijkl}$ are electrostriction tensor components, $\sigma_{ij}$ are elastic stresses, and $\mu$ is the coefficient of thermopolarization effect [86]; $F_{ijkl}$ are flexoelectric tensor coefficients.

The important aspect of the Eq. (2b) is the presence of the thermopolarization coupling, $\mu$. Here, the coefficient $\mu$ is the diagonal component of the second rank tensor $\mu_{ij}$, which value can be estimated as proportional to the convolution of the flexoelectric tensor $F_{ijkl}$ and linear thermal expansion tensor $\beta_{ij}$, namely $\mu_{ij} \cong F_{ijkl}\beta_{kl}$ [87]. Note that the thermopolarization effect is omnipresent, meaning that it exists for arbitrary symmetry of the studied material [86], but its numerical values are poorly known [75]. We further restrict the analysis to the transversally isotropic thermal expansion tensor $\beta_{ij} = \delta_{ij}\beta_{ii}$ with $\beta_{11} = \beta_{22} \neq \beta_{33}$ ($\delta_{ij}$ is the Kroneker symbol).

The electric boundary conditions are the fixed potential $V$ at the tip-ferroelectric contact area, $\phi^{(in)}\big|_{S_t} = V$, electric potential and displacement continuity at the FE surface, $\left(\phi^{(out)} - \phi^{(in)}\right)\big|_{z=0} = 0$ and $\left(-\varepsilon_0\varepsilon_b\frac{\partial\phi^{(in)}}{\partial z} + P_3 + \varepsilon_0\varepsilon_e\frac{\partial\phi^{(out)}}{\partial z}\right)\big|_{z=0} = 0$, and potential vanishing at bottom electrode, $\phi^{(in)}\big|_{z=h} = 0$, (or at the infinity at $h \to \infty$). The potential $\phi^{(out)}$ obeys the Laplace equation outside the FE. The so-called "natural conditions" are valid for the polarization at the FE surfaces, $\left(\frac{\partial P_3}{\partial z}\right)\big|_{z=0,h} = 0$.

Elastic stresses $\sigma_{ij}$ and strains $u_{ij}$ are calculated in a self-consistent way from elastic equations in the continuum media approach. The elastic equations of state follow from the variation of the LGD free energy with respect to elastic stresses:

$$u_{ij} = s_{ijkl}\sigma_{kl} + \beta_{ij}(T - T_0) + F_{ijkl}\frac{\partial P_l}{\partial x_k} + Q_{ijkl}P_kP_l, \quad 0 \leq z \leq h. \tag{3a}$$

The strain tensor components are related to the displacement components $U_i$ in a conventional way, $u_{ij} = (\partial U_i/\partial x_j + \partial U_j/\partial x_i)/2$.

Note that the linearization of electrostriction terms with respect to electric field gives the piezoelectric contribution in a FE phase, namely using the expression for polarization, $P_k = P_k^S + \chi_{kn}E_n$, the electrostriction contribution $Q_{ijkl}P_kP_l \cong Q_{ijkl}P_k^SP_l^S + d_{ijm}E_m + Q_{ijkl}\chi_{lm}E_m\chi_{kn}E_n$, where $d_{ijm} = 2Q_{ijkl}P_k^S\chi_{lm}$ is a piezoelectric tensor expressed via the electrostriction $Q_{ijkl}$, spontaneous polarization $P_k^S$ and dielectric susceptibility $\chi_{lm}$ tensors.

Equations (3a) should be solved along with equations of mechanical equilibrium

$$\partial\sigma_{ij}(\mathbf{x})/\partial x_i = 0, \tag{3b}$$

and compatibility equations, $e_{ikl}e_{jmn}\partial^2 u_{ln}(\mathbf{x})/\partial x_k\partial x_m = 0$, which are equivalent to the continuity of $U_i$ [88]. The boundary conditions for elastic stresses $\sigma_{ij}$ and displacement components $U_i$ at the FE surfaces are the absence of normal stress at the free top surface, $\sigma_{i3}|_{z=0} = 0$, and zero elastic displacement due to complete clamping at the substrate electrode, $U_i|_{z=h} = 0$. Here we assume that the deformation of the top surface is small, otherwise we need to apply the boundary condition at the (unknown) deformed boundary.

To complement analytical derivations, finite element modeling (**FEM**) is performed in a COMSOL@MultiPhysics software, using electrostatics, solid mechanics, and general math (PDE toolbox) modules. To avoid numerical artefacts, the temperature and voltage distribution at the ferroelectric film surface is chosen as Gaussian-like with a dispersion $R$.

As representative model systems, here we explore different types of uniaxial ferroelectrics: a "soft" ferroelectric $Sn_2P_2S_6$ (**SPS**) with a relatively low bulk Curie temperature $T_C$=337 K and small coercive field, and a "hard" ferroelectric-pyroelectric $LiNbO_3$ (**LNO**) with a high $T_C$ = 1477 K and ultra-high coercive field. We performed a quasi-2D simulations for a 100-nm thick SPS and LNO layers. The corresponding LGD free energy coefficients and other material parameters are listed in **Table I.** The free energy dependence on the polarization $P$ and temperature $T$ is shown in **Fig. 1(d)** for $Sn_2P_2S_6$ and **1(e)** for $LiNbO_3$, respectively.

**Table I.** The parameters for bulk ferroelectrics $Sn_2P_2S_6$ and $LiNbO_3$

| Parameter | Dimension | Values for $Sn_2P_2S_6$ collected from Refs. [89, 90, 91] | Values for $LiNbO_3$ collected from Refs.[92, 93, 94, 95, 96, 97] |
|---|---|---|---|
| $\varepsilon_b$ | 1 | 7 [*] | 4.6 [92] |
| $\alpha_T$ | m/F | $1.44\times10^6$ | $1.569\times10^6$ [93] |
| $T_C$ | K | 337 | 1477 Ref. [94] |
| $\beta$ | $C^{-4}\cdot m^5 J$ | $9.40\times10^8$ | $2.31\times10^9$ [***] |
| $\gamma$ | $C^{-6}\cdot m^9 J$ | $5.11\times10^{10}$ | $1.76\times10^9$ [***] |
| $g_{ij}$ | $m^3/F$ | $g_{11}=5.0\times10^{-10}$ [**] $g_{44}=2.0\times10^{-10}$ | $g_{44}=7.96\times10^{-11}$ [95] |
| $s_{ij}$ | 1/Pa | $s_{11}=4.1\times10^{-12}$, $s_{12}=-1.2\times10^{-12}$, $s_{44}=5.0\times10^{-12}$ | $s_{11}=5.78\times10^{-12}$, $s_{12}=-1.01\times10^{-12}$, $s_{13}=-1.47\times10^{-12}$, $s_{33}=+5.02\times10^{-12}$, $s_{14}=-1.02\times10^{-12}$, $s_{44}=17.10\times10^{-12}$ [96] |
| $Q_{ij}$ | $m^4/C^2$ | $Q_{11}$=0.22, $Q_{12}$=0.12 [****] | $Q_{33}$= +0.016, $Q_{13}$= −0.003 |
| $\mu$ | V/K | $6.0\times10^{-5}$ [*****] | $6.0\times10^{-5}$ [*****] |
| $F_{ij}$ | $m^3/C$ | $F_{11}=1.0\times10^{-11}$, $F_{12}=0.9\times10^{-11}$, $F_{44}=3\times10^{-11}$ | $F_{11}=1.0\times10^{-11}$, $F_{12}=0.9\times10^{-11}$, $F_{44}=3\times10^{-11}$ |
| $\beta_{ij}$ | 1/K | $\beta_{11}=\beta_{22}=4\times10^{-5}$, $\beta_{33}=9\times10^{-6}$ | $\beta_{11}=14.4\times10^{-6}$, $\beta_{22}=15.9\times10^{-6}$, $\beta_{33}=7.5\times10^{-6}$ [97] |

[*] estimated from a refraction index value

** the order of magnitude is estimated from the uncharged domain wall width [89, 90].

*** The estimation is based on the values of the spontaneous polarization and permittivity at room temperature

**** Estimation of electrostriction is based on thermal expansion data from Say et al.[91]

***** Estimated as a convolution of the flexoelectric and thermal expansion tensors, and the numbers order is the same as in Ref.[75]

From Eqs.(3a), the local elastic strain (and hence PFM response) has several contributions coming from the thermal expansion [98], flexoelectric effect [99], and from electrostriction that includes the piezoelectric effect [100] and the thermopolarization effect [86]. The flexoelectric and thermopolarization contributions are universal, while the piezoelectric contribution is symmetry-sensitive being dominant in the ferroelectric phase without inversion symmetry. In the analysis below, we also neglect the chemical pressure (Vegard contributions) that underpin signal formation mechanisms in electrochemical strain microscopy [99, 101, 102], and temperature-induced shifts of electrochemical equilibrium at the free surfaces [103, 104, 105], and defer these mechanisms to future studies.

## III. LOCAL HEATING-INDUCED POLARIZATION CHANGES AND STRAINS AT ZERO TIP VOLTAGE

In this section, we analyze the phenomena emerging under the local heating. Note that the basic insight into the relevant phenomena can be derived from joint consideration of the temperature dependence of polarization and long-range nature of depolarization fields in ferroelectrics. Namely, local heating of the ferroelectric surface necessarily reduces the polarization below the tip, resulting in the polarization gradient within the material. The polarization gradient is in turn associated with the polarization bound charge, that in turn can be minimized via the penetration of the region with reduced polarization inside the material, or clamping of polarization below the tip to higher (relative to equilibrium) values.

To gain insight in these phenomena, we consider a thick FE layer placed under the heated tip when the voltage applied between the tip and the bottom electrode is zero, $V = 0$. The layer was homogeneously polarized before the heating. Spatial distributions of the temperature excess $T$, polarization $P_3$ and vertical displacement $U_3$ of the FE layer are shown in **Fig. 2**. XZ cross-sections are calculated by FEM for two values of the tip "overheating" on $\Delta T = 50$ K for SPS (left column) and $\Delta T = 500$ K for LNO (right column), tip-surface contact radius $R = 10$ nm and zero applied voltage, $V = 0$. The heated region has a semi-spherical profile [compare **Fig. 2(a)** with **2(d)**].

At zero voltage, the heating-induced changes of $P_3$ is primary caused by the temperature changes of the coefficient $\alpha_T(T(\vec{r}) - T_C)$, and also by the thermopolarization and flexoelectric effects [see the right-hand side of Eq.(2b)]. Based on numerical estimates for materials explored here, the flexoelectric contribution is small in comparison with the thermopolarization contribution for initially homogeneously polarized FE (see **Appendix B**).

A small overheating on $\Delta T = 50$ K significantly decreases the ferroelectric polarization $P_3$ in the overheated region of SPS [see **Fig.2(b)**]. The region of reduced polarization growths through the SPS layer depth in order to minimize the strong depolarization field produced by a charged domain wall [106, 107]. The polarization behavior is relatively easy to rationalize – the thermal field is localized below the probe, but the ferroelectric polarization cannot form z-gradients due to the strong depolarization filed. As a result, the area with reduced polarization extends far beyond the heated region and induced corresponding changes of the elastic fields in the same region. Note that this effect is dual, and polarization below the tip is also clamped by the surrounding material.

In comparison, for LNO even high overheating on $\Delta T = 500$ K neither induces the polarization decrease nor the nanodomain formation in the overheated region of LNO [see **Fig.2(c)**]. Only the bulb-like region of slightly suppressed polarization appears in the case. The polarization behavior is explained by the fact that overheating on 500 K is still very far from high $T_C$ of LNO, and so the FE remains insensitive to the overheating.

Further shown are the corresponding changes of the vertical displacement $U_3$ originated from the thermal expansion (i.e., from the thermoelastic effect), electrostriction and flexoelectric effects [see the right-hand side of Eq.(3a)]. Since the flexoelectric contribution appeared negligibly small, $U_3$ profiles are controlled by the thermoelastic and electrostriction contributions. The region of the heating-induced $U_3$ is much wider and much more diffuse than the region of temperature excess for both SPS and LNO layers [compare **Fig. 2(a)** with **2(c)**, and **Fig. 2(d)** with **2(f)**, respectively]. Both x- and z-profiles of $U_3$ "falls down" as a whole in the region with radius $r \gg R$; and reveal a diffuse maximum in the region with radius $r \cong 2R$.

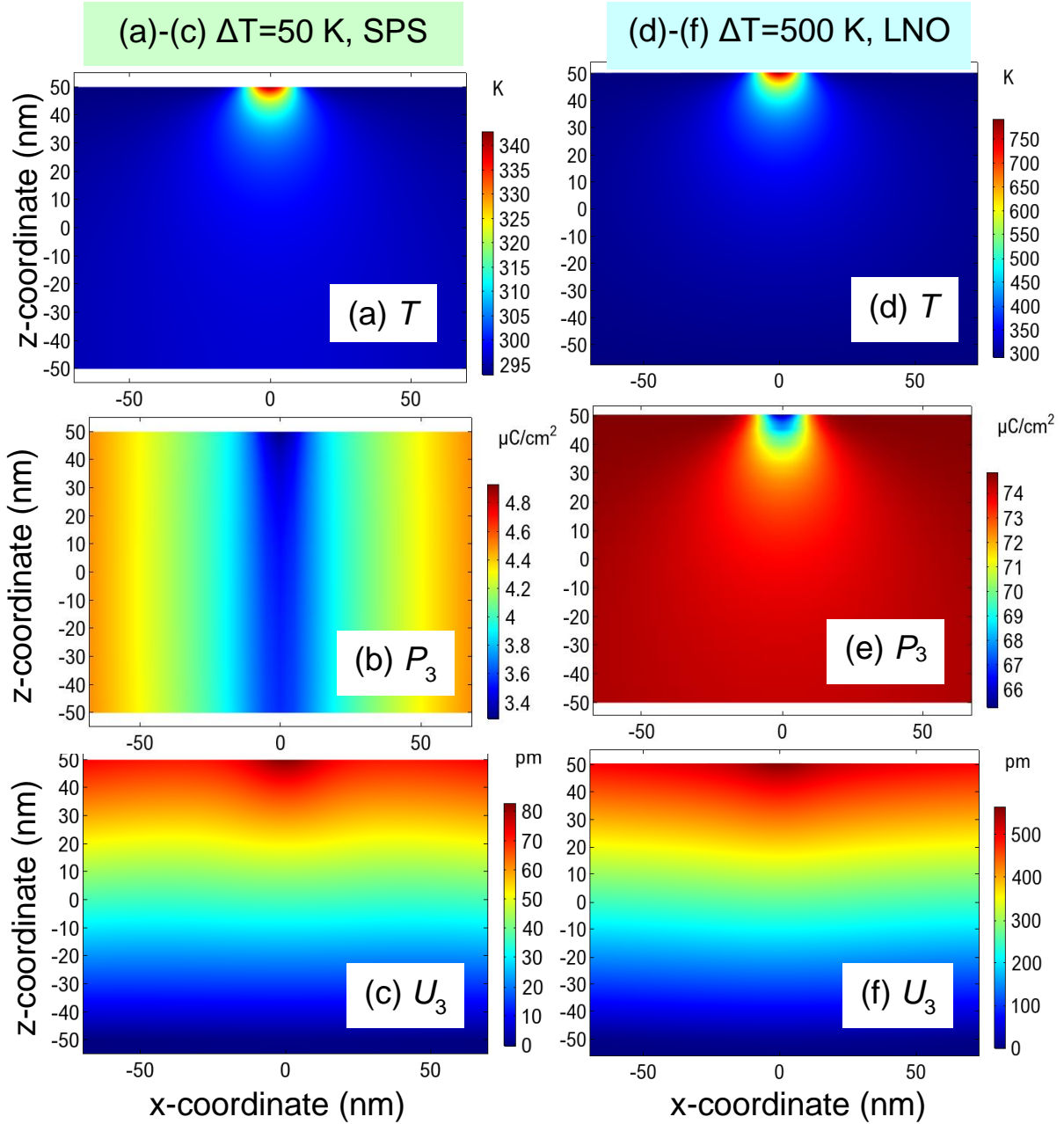

**FIG. 2**. Spatial distributions of the temperature $T$ (**a, d**), polarization $P_3$ (**b, e**) and vertical displacement $U_3$ (**c, f**) of the FE layer calculated by FEM for two values of the tip overheating $\Delta T = 50$ K for SPS parameters (**a-c**) and 500 K for LNO parameters (**d-f**); tip-surface contact radius $R = 10$ nm, and $T_0 = 293$ K. X-Z cross-sections are shown. Applied voltage is absent, $V = 0$, the FE was homogeneously polarized before the heating.

## IV. LOCAL HEATING-INDUCED POLARIZATION CHANGES, STRAINS, AND ELECTROMECHANICAL RESPONSE AT NONZERO TIP VOLTAGES

In this section we consider a thick FE layer placed under the heated tip when the voltage is applied between the tip and the bottom electrode. The layer was homogeneously polarized before probing.

### A. FEM results for polarization and elastic displacement

We further explore the joint effect of probe heating and bias in the tPFM experiments. Typical spatial distributions of the polarization $P_3$ and vertical displacement $U_3$ of the FE calculated by FEM for small and high values of applied voltage $V = \pm 0.1$ V, $\pm 1$ V and $\pm 10$ V and tip overheating $\Delta T = 50$ K for SPS and 500 K for LNO ferroelectrics are shown in **Fig. 3** and **4**, respectively.

For a chosen "up" direction of spontaneous polarization [shown in **Figs. 1(b)-(c)**], the negative voltage increases the spontaneous polarization under the heated tip, but quantitatively the polarization enhancement is different for SPS [**Fig. 3(a)**] and LNO [**Fig. 4(a)**]. Specifically, for SPS the polarization enhancement occurs in the stripe region that penetrates through the layer depth; and the stripe is surrounded by the region of suppressed polarization [**Fig. 3(a)**]. For LNO the polarization enhancement occurs in a small semi-ellipsoidal region that does not penetrates into the layer [**Fig. 4(a)**]. Also, note the unusual structure of the tip-induced polarization suppression for both small and higher negative voltages, shown by dark-blue satellites in **Fig. 3(a)** and **4(e)**.

Positive voltages decrease the polarization. For sufficiently high magnitude $V > V_{th}$, the bias applied to the tip can reverse local polarization and induce the nanodomain [10, 11]. The threshold voltage $V_{th}$ is estimated to be very low for an SPS film (less than 10 mV), and rather high for LNO – more than 5 V nanodomain [compare **Fig. 3(b)** and **4(f)**]. Note that these estimates strongly depend on tip radius of curvature and potential drop at the tip-surface junction (dead layer effect, [10, 11]). The nanodomain breakdown through the layer immediately occurs in SPS at $V > V_{th}$ [**Fig. 3(b)**]. In LNO the spike-like nanodomain nucleus occurs at high voltage $V_{th} \cong 5$ V, and its breakdown happens at significantly higher voltages [**Fig. 4(f)**]. The structure is conditioned by the system tendency to minimize the depolarization field energy that appear near any sort of polarization gradient with nonzero divergency.

At small voltages, the displacement maps are almost insensitive to the direction of SPS polarization under the tip [see **Fig. 3(c)** and **Fig. 3(d)**], and the difference becomes even smaller for LNO with voltage increase [see **Fig. 4(c), 4(d), 4(g)** and **4(h)**]. This insensibility is caused by the quadratic electrostriction effect. Only the voltage derivative (i.e., piezoelectric contribution) can be sensitive. Spatial distributions of the temperature $T$ are voltage-independent, so they are the same as shown in **Fig. 2(a)** and **2(d).**

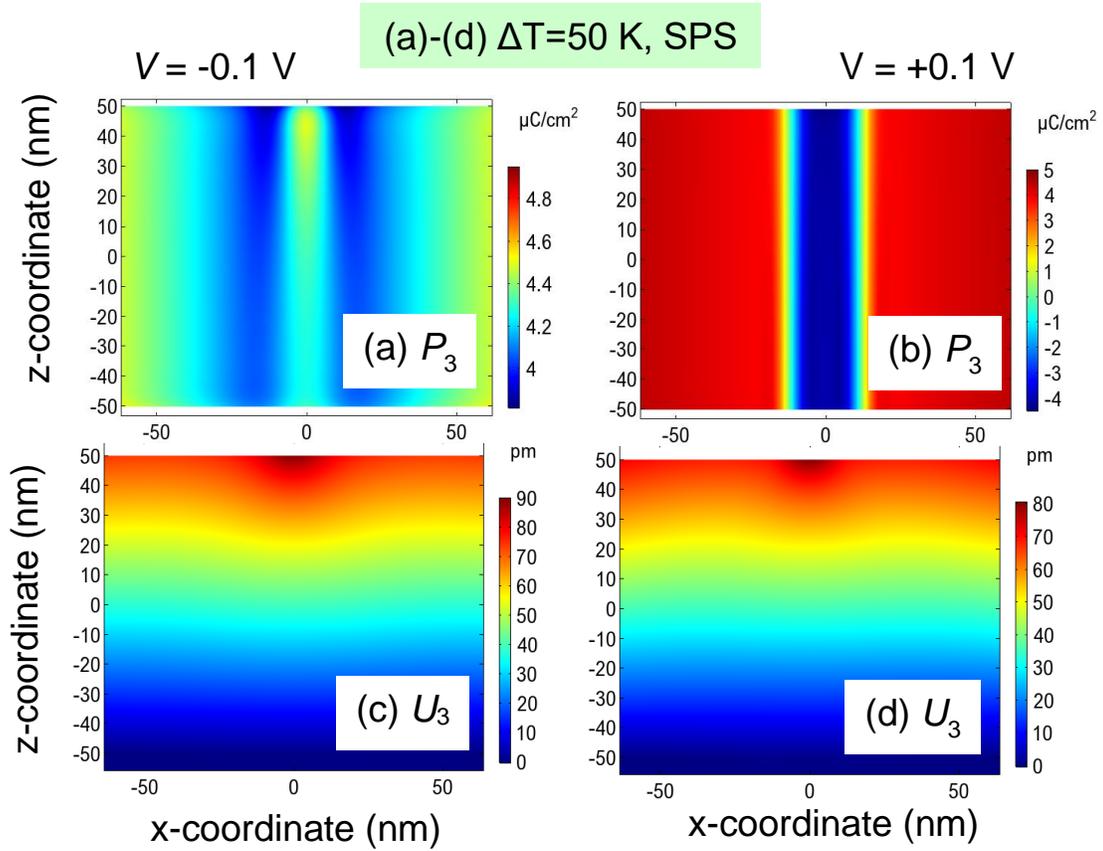

**FIG. 3**. Spatial distributions (*xz* cross-sections) of the polarization $P_3$ **(a, b)** and vertical displacement $U_3$ **(c, d)** of a thick SPS layer calculated for applied voltage $V = -0.1$ V **(a, c)**, $V = +0.1$ V **(b, d)**, tip-surface contact radius $R = 10$ nm, tip overheating $\Delta T = 50$ K, and $T_0 = 293$ K. Before heating the SPS layer was homogeneously polarized. Material parameters are listed in **Table I.**

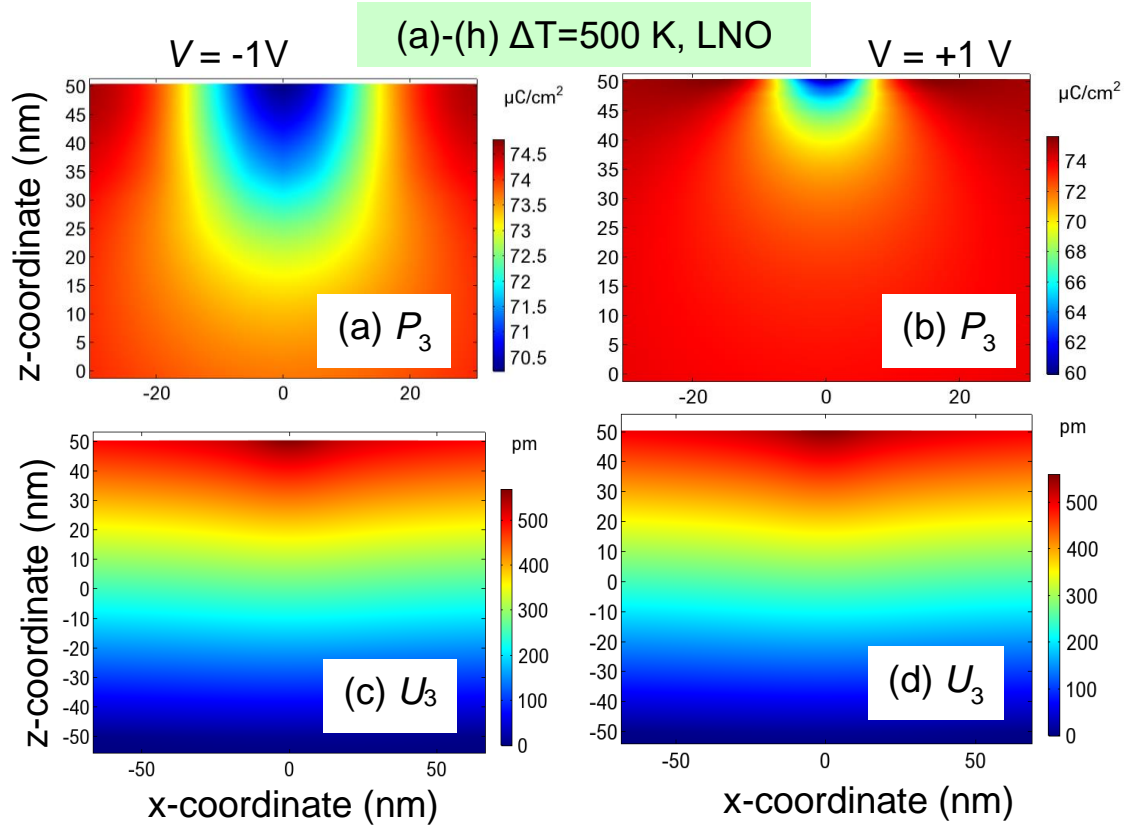
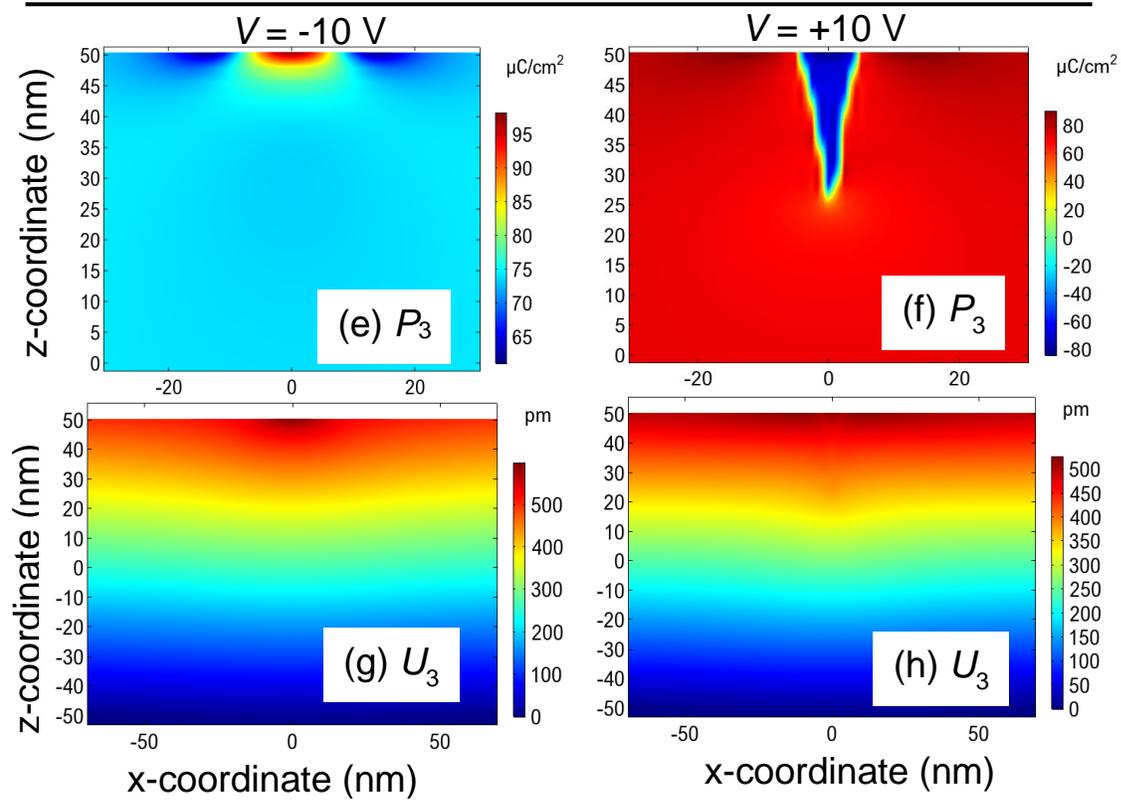

(a)-(h) ΔT=500 K, LNO

**FIG. 4**. Spatial distributions ($xz$ cross-sections) of the polarization $P_3$ **(a, b, e, f)** and vertical displacement $U_3$ **(c, d, g, h)** of a thick LNO layer calculated for applied voltage $V = -1$ V **(a, c)**, $V = +1$ V **(b, d)**, $V = -10$ V **(e, g),** and $V = +10$ V **(f, h)** for the tip-surface contact radius $R = 10$ nm, tip overheating $\Delta T = 500$ K, and $T_0 = 293$ K. Before heating the LNO layer was homogeneously polarized. Material parameters are listed in **Table I.**

**Figures 3-4** are calculated for nonzero flexoelectric coefficients, $F_{ij} \neq 0$, which are listed in **Table I**, and whose order of magnitude are the same as for other ferroelectrics [108]. It is seen from the **Fig. A1**, that the flexoelectric coupling does not affect the displacement distribution significantly. In fact, the flexoelectric effect contribution is negligibly small even at the diffuse domain walls shown in **Fig. 3-4** and **A1**. However, this observation can be readily rationalized since for heating of $\Delta T > 5$ K and nonzero voltages $|V| > 0.05$ V the piezoelectric and electrostriction contributions strongly dominate over the flexoelectric contribution, and as well as over the thermopolarization contribution.

The profiles of polarization $P_3$ and vertical displacement $U_3$ at the FE surface calculated for tip overheated at $\Delta T = 50$ K and $\Delta T = 500$ K, positive, zero and negative voltages $V$ are shown in **Figs. 5(a)-5(b)** for SPS layer and in **Figs. 5(c)-5(d)** for LNO layer, respectively**.** Black solid curves in **Figs. 5(a)** and **5(c)**, calculated for zero voltage $V = 0$, show the changes of the $P_3$ surface profiles induced by the thermopolarization effect, which role is little more pronounced for SPS in comparison with LNO. The $P_3$ profiles calculated for nonzero voltages [colored curves in **Figs. 5(a)** and **5(c)**] are smoother for SPS, where the ferroelectric polarization is enhanced or reversed by the biased heated tip at much lower voltages (~0.1 V) than for LNO (~ 10 V). Note that SPS is very a "soft" ferroelectric for tPFM in comparison with a "hard" LNO. Interestingly, that the field-induced polarization conserves regardless the heating in LNO up to very high temperatures (more than 1000 K). For a hard ferroelectric the tip overheating well above $T_C$ (on more than 100-500 K) is required to induce a local transition to the paraelectric phase, but such strong overheating can rather melt the ferroelectric.

The temperature- and voltage-induced surface profiles of $U_3$, which are caused by the thermoelastic and electrostriction effects, look very different for SPS and LNO, compare **Figs. 5(b)** and **5(d)**. For SPS the $U_3$ profiles have a maximum at the center for both negative and positive voltages. The maxima height depends on the tip voltage in a very specific way: it is the smallest for 0.1 V, becomes biggest for 0, -0.1V, +0.5V, -0.5V, +1V and highest for -1V. The "alternating" sequence is related with the interplay of elastic responses from the overheated nanoregion (or reversed nanodomain) and colder FE surrounding. For LNO the $U_3$ profiles have a single central maximum for negative and relatively small positive voltages, which splits in 2 or 3 maxima for higher voltages. The maxima height depends on the tip voltage in a monotonic way: it is the smallest for 10 V, becomes biggest for 5 V, 1 V, 0, -1 V, -5 V and highest for -10 V.

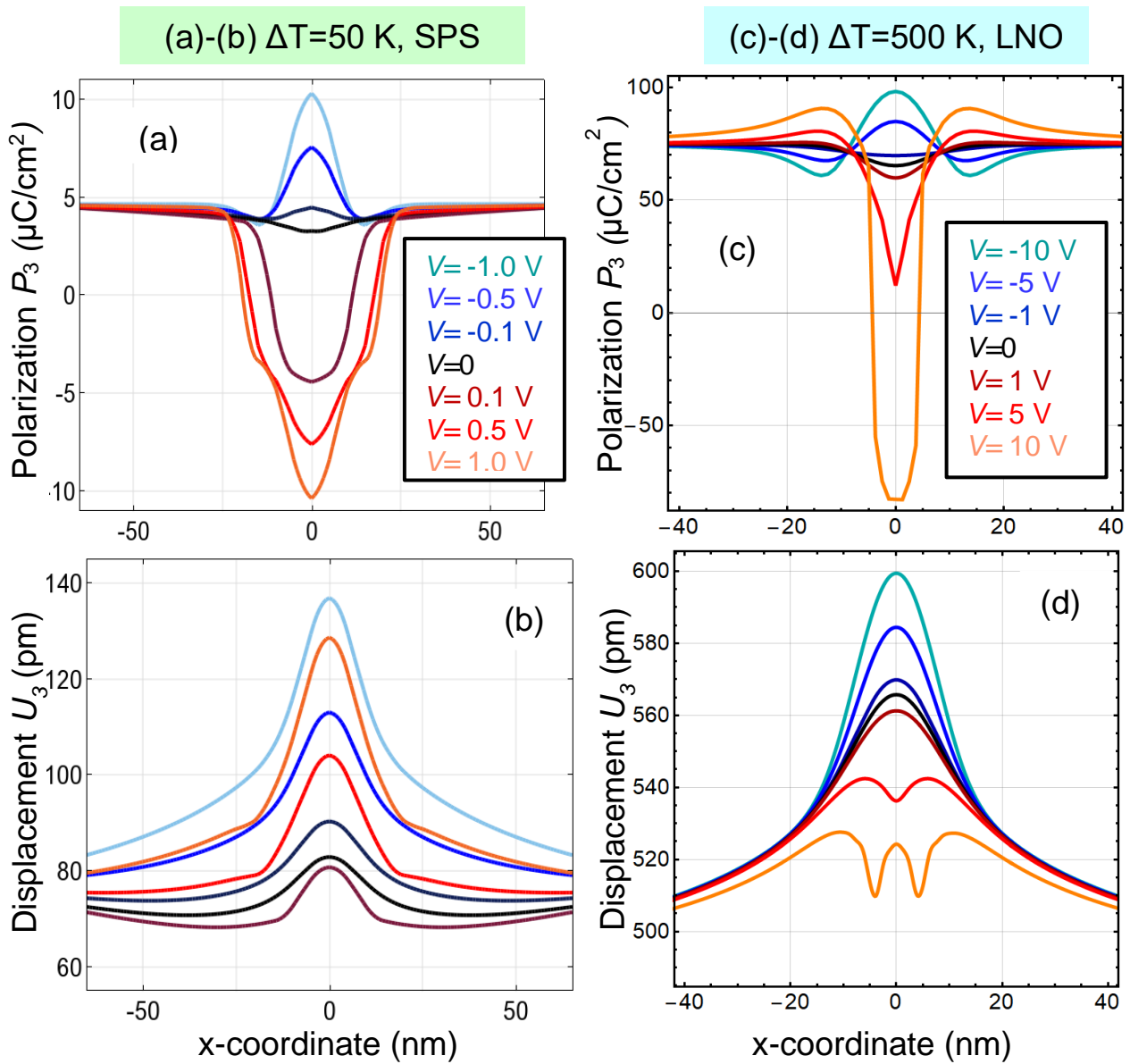

**FIG. 5.** Polarization (**a, c**) and vertical displacement $U_3$ (**b, d**) profiles at the FE surface calculated for the tip overheating $\Delta T = 50$ K and SPS parameters (**a, b**); and $\Delta T = 500$ K and LNO parameters (**c, d**). Tip

voltage $V$ varyes from -1V to +1V for SPS, and from -10V to +10V for LNO (see legends). Before heating the FE was homogeneously polarized. Other parameters are the same as in **Fig. 4**.

## B. Temperature dependences of polarization, elastic displacement and local electromechanical response

The temperature dependences of the polarization $P_3$ and vertical displacement $U_3$ (in pm) calculated under the headed tip (centered at x=0) are shown in **Figs. 6(a)-6(b)** for SPS layer and in **Fig. 6(c)-6(d)** for LNO layer, respectively. Note that the temperature dependences look very different for SPS and LNO; and they have very different sensitivity to the temperature and applied voltage.

First, we discuss the temperature dependence of polarization, displacement, and electromechanical response for a soft SPS. Black solid curves in **Figs. 6(a)**, calculated for zero voltage, show the temperature dependence of the $P_3$ induced by the thermopolarization effect. Black solid curves in **Figs. 6(b)**, also calculated for $V = 0$, show the temperature dependences of $U_3$ induced by the thermoelastic effect and electrostriction. Both these dependences have a feature at about $\Delta T_{cr} = 80$ K, where the ferroelectric polarization is destroyed under the heated tip, indicating on the local temperature-induced transition to a paraelectric phase. Dashed black curves, calculated for very small voltages $V = \pm 10$ mV, are relatively close to the solid black curves for polarization and displacement. Dark red, red and orange curves calculated for positive voltages $V = (0.1 - 1)$ V are mostly linear, except for the very thin temperature region of polarization reversal at $V = 0.1$ V. For higher voltages the field-induced polarization conserves regardless the heating. Dark blue, blue and teal curves calculated for negative voltages $V = -(0.1 - 1)$V are quasi-linear for the same reasons. The displacements for positive and negative voltages become closer with temperature increase [see **Fig. 6(b)**].

The situation for a hard FE – LNO differs strongly from SPS. Black solid curves in **Figs. 6(c)**, calculated for $V = 0$, show the linear temperature dependence of the $P_3$ mostly induced by the linear thermal expansion and also by the thermopolarization effect. The tip overheating well above $T_C$ (on more than 100-500 K) is required to induce a local phase transition at $V = 0$, but such strong overheating can rather destroy the ferroelectric. For high voltages (both positive or negative) the field-induced polarization conserves regardless the heating. At the same time, the nucleation of a spike-like nanodomain occurs at high voltages (~5 – 10 V). Displacement curves, calculated for both negative, zero and positive voltages, are linear due to the dominant contribution of the linear thermal expansion [see all curves in **Figs. 6(d)**], regardless the polarization reversal occurs at $V = 5$ V and 700 K [see the red curve in **Figs. 6(c)**]. The displacement curves for positive and negative voltages remain parallel with temperature increase as anticipated for the linear thermal expansion mechanism.

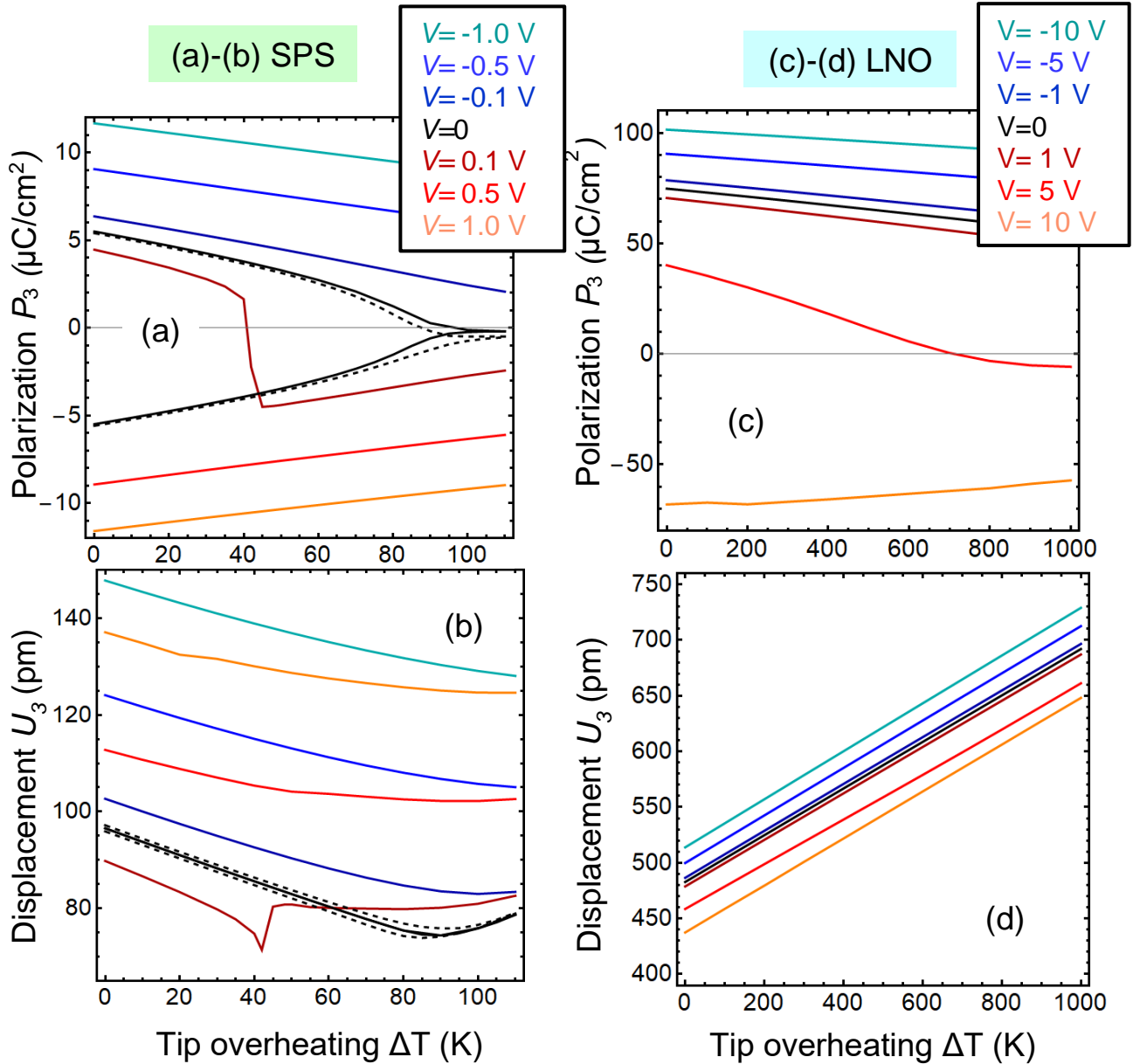

**FIG. 6.** Surface polarization $P_3$ (**a, c**) and vertical displacement $U_3$ (**b, d**) versus $\Delta T$ calculated under the headed tip (centered at x=0). Solid curves are calculated for different voltages $V$ varying from -1V to +1V

for SPS, and from -10V to +10V for LNO (see legends). Dashed black curves in plots (a)-(b) correspond to $V = \pm 10$ mV. Before heating the FE was homogeneously polarized. Other parameters are the same as in **Fig. 4.**

The temperature dependence of the effective local electromechanical response $d_{33}^{eff}$, which determines the PFM response, can be calculated from the expression

$$d_{33}^{eff}(x,V) = \frac{dU_3(x,V)}{dV} \approx \frac{U_3(x,V+\delta V) - U_3(x,V-\delta V)}{2\delta V}, \quad (4)$$

where $\delta V$ must be very small (e.g., note more than several mV). FEM results are shown in **Fig. 7**

Note that the temperature dependences look very different for SPS [**Fig. 7(a)**] and LNO [**Fig. 7(b)**]; and for SPS we can expect more strong dependence on the temperature and applied voltage. In particular, the temperature dependence of $d_{33}^{eff}$ for SPS has a diffuse maximum (or break) at about $\Delta T_{cr} = 85$ K indicating on the temperature-induced local paraelectric transition under the heated tip. The transition is absent for LNO for all voltages, except for 5 V. Since SPS surface displacements for positive and negative voltages become rather close with temperature increase [**Fig. 6(a)**], their voltage derivatives are also close, but have different signs and demonstrate a noticeable break at $\Delta T_{cr}$ for $V = \pm 0.1$ V and 0.1 V [see dashed curves in **Fig. 7(a)**].

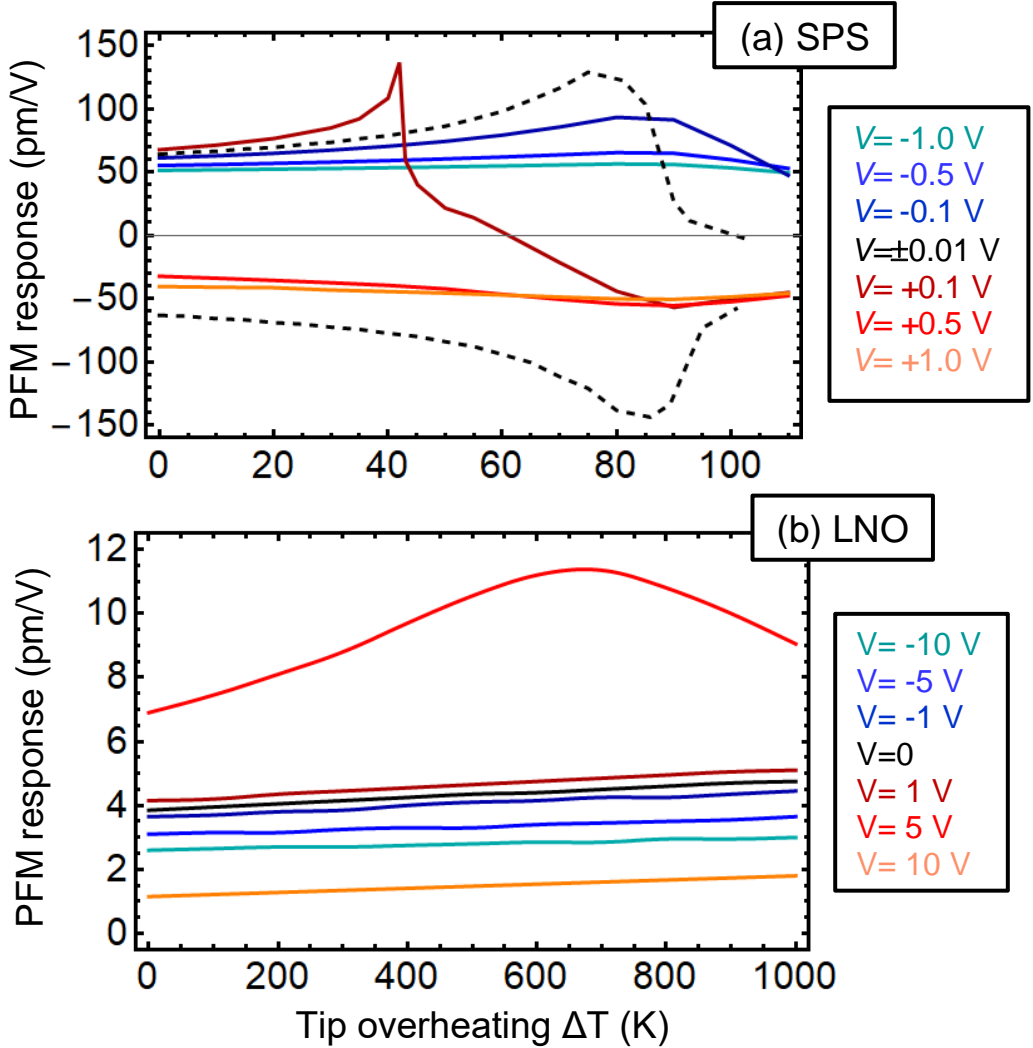

**FIG. 7.** Effective response $d_{33}^{eff}$ versus $\Delta T$ calculated under the headed tip (centered at x=0). Different curves are calculated for different voltages $V$ varying from -1V to +1V for SPS, and from -10V to +10V for LNO (see legends). Before heating the FE was homogeneously polarized. Other parameters are the same as in **Fig. 4.**

### C. Approximate analytical description of the local electromechanical response

Finally, we analyze the image formation mechanism in the thermal PFM can be considered within a classical continuous medium approach, which offers possibilities for analytical calculations in various ferroics within e.g., decoupling approximation [47, 109, 110] well-elaborated for classical ferroelectrics [100]. Here, PFM response has several contributions coming from the thermal expansion, flexoelectric, and electrostriction including thermopolarization and piezoelectric effects [see Eq.(3a)]. At that the piezoelectric effect is a derivative of the electrostriction term [see comments to Eq.(3a)]. Below we analyze and list approximate expressions for all these contributions.

The thermoelastic contribution to the surface displacement comes from the inhomogeneous thermal expansion of the FE caused by the contact with a heated tip, and the expansion is proportional to the specific integral convolution of the material thermal expansion tensor $\beta_{ij}$ with the temperature variation. For the temperature excess given by a uniformly heated disk, the vertical displacement of the FE surface caused by the thermoelastic effect is given by expression:

$$U_3^{TE}(r) = \Delta T \frac{\beta_{11}(1+4\nu)+3\beta_{33}}{2\pi} arcsin\left(\frac{2R}{|r-R|+|r+R|}\right), \quad (5)$$

where the factor $\nu$ is the Poisson ratio. Expression (5) is derived in **Appendix A**. As it was expected, the magnitude of the thermoelastic contribution to PFM response is proportional to the tip temperature variation $\Delta T$ and thermal expansion coefficients combination $\beta_{11}(1+4\nu)+3\beta_{33}$. The thermoelastic effect is voltage-independent, and the spatial region of its maximal values is $r \leq R$, so the contribution is not responsible for the wide region $r \gg R$ of $U_3$ changes. Since the thermoelastic effect is voltage-independent, it does not contribute to the PFM signal detected via the lock-in or band-excitation [111] detection.

In decoupling approximation, the electrostriction contribution to the surface displacement is proportional to the integral convolution of the local term $Q_{ijkl}P_k(\mathbf{r})P_l(\mathbf{r})$ with elastic Green function (see **Appendix A** and Ref. [100]). The displacement profile complexly depends on the temperature profile due to the integration. The following Pade-approximation can be used for semi-quantitative analysis of the FEM data:

$$U_3^{EL}(r) \sim Q \frac{2\pi R^2 h}{\sqrt{R^2+\varepsilon r^2}} P_3^2(r) \quad (6a)$$

Here $Q$ is the combination of electrostriction coefficients and elastic constants, $\varepsilon$ is the fitting parameter varying in the range $0 \leq \varepsilon \ll 1$. Hence the denominator $\sqrt{R^2+\varepsilon r^2}$ has a diffuse maximum in a region $r \gg R$. In order to obtain a simple expression for $P_3^2(r)$, let us regard $\beta > 0$ and neglect $\gamma$ in Eq.(2b). The amplitude of the FE polarization is proportional to,

$$P_3(r) \sim \mu \frac{dT}{dZ} + \chi V + \sqrt{\frac{\alpha_T}{\beta}} \begin{cases} \sqrt{T_C - T_0 - \epsilon \Delta T}, & T_C - T_0 \geq \epsilon \Delta T, \\ 0, & T_C - T_0 < \epsilon \Delta T, \end{cases} \quad (6b)$$

where the fitting parameters are $\mu$, $\chi$ and $\epsilon$; at that $0 \leq \epsilon < 1$. The first term originated from the thermopolarization effect, the second term is proportional to the tip voltage and dielectric susceptibility $\chi$, and the third term is the spontaneous polarization.

Using a disk-contact model of the tip-surface contact, the piezoelectric contribution to the vertical PFM response from a homogeneously polarized FE region is [83]:

$$d_{33}^{eff} = \frac{dU_3}{dV} \approx \left(\frac{1}{4}+v\right)d_{31} + \frac{3}{4}d_{33} + \frac{d_{15}}{4}, \tag{7a}$$

The expression (7a) is valid for a "cold" PFM response. For the heated tip the effective piezoelectric coefficients $d_{ij}$ are dependent on the distance from the tip apex $r$. In the sense of the local approximation:

$$d_{ij}(r) \cong d_{ij3}^0 \chi P_3(r), \tag{7b}$$

Expressions (7) are valid is the case of very smooth polarization changes under the tip and small heating, e.g., at $\epsilon \Delta T \ll T_C - T_0$.

## V. CONCLUSION

The signal formation mechanisms of the tPFM on the semi-infinite ferroelectric surface is analyzed. Here we explored the solution of a thermo-elastic-electric problem fully coupled with Landau-Ginzburg-Devonshire description of ferroic properties on two different types of uniaxial ferroelectrics: a "soft" ferroelectric SPS with a relatively low bulk Curie temperature $T_C$<350 K and relatively small coercive field, and a "hard" ferroelectric-pyroelectric LNO with a high $T_C$ >1300 K and ultra-high coercive field.

The temperature-induced polarization redistribution and local electromechanical response occurring in these two ferroelectrics under the heated PFM tip strongly depend on the material parameters and, surprisingly, reveal very different sensitivities to the temperature $T$ and tip voltage $V$. Specifically, for a soft ferroelectric, the tip overheating by 30ºC above $T_C$ leads to the local paraelectric transition in the nanoscale region at $V = 0$. The tip voltage-induced nucleation of a nanodomain and its subsequent breakdown through the film depth occur at very low voltages V~ (10 – 100) mV. The contribution of the thermopolarization effect to the local electromechanical response of the soft ferroelectric appears very important. For a hard ferroelectric, the tip overheating well above $T_C$ by more than 100-500 K is required to induce a local paraelectric transition at $V = 0$, but such strong overheating can rather melt the ferroelectric. The nucleation of a spike-like nanodomain occurs at high tip voltages ~(5 – 10) V. The contribution of the thermopolarization effect to the local electromechanical response of the hard ferroelectric is less significant than that for a soft ferroelectric. As anticipated, the tPFM response is a little sensitive to the flexoelectric effect in both types of ferroelectrics, and the response is determined by the piezoelectric and electrostriction contributions.

Overall, tPFM opens the pathway for probing local temperature induced phase transitions in ferroics, exploring the temperature dependence of polarization dynamics in ferroelectrics, and potentially discovering coupled phenomena driven by strong temperature- and field gradients. The tPFM is a promising tool for the exploration of the temperature-induced nanoscale phase transitions in ferroics, such as ferroelectrics, antiferroelectrics, quantum paraelectrics and related materials.

**Authors' contribution.** A.N.M. and S.V.K. stated the problem, interpreted theoretical results and wrote the manuscript draft. A.N.M. propose the mathematical model and performed analytical calculations, E.A.E. performed FEM. K.K. worked on the manuscript improvement and results interpretation.

**Funding.** This effort (K.K., S.V.K.) was supported by the center for 3D Ferroelectric Microelectronics (3DFeM), an Energy Frontier Research Center funded by the U.S. Department of Energy (DOE), Office of Science, Basic Energy Sciences under Award Number DE-SC0021118, and the Oak Ridge National Laboratory's Center for Nanophase Materials Sciences (CNMS), a U.S. Department of Energy, Office of Science User Facility. A.N.M. work has been supported by the National Research Fund of Ukraine (project "Low-dimensional graphene-like transition metal dichalcogenides with controllable polar and electronic properties for advanced nanoelectronics and biomedical applications", grant application 2020.02/0027).

# SUPPLEMENT

## Appendix A. Decoupled thermo-elastic problem. Approximate analytical solution

Let us regard that a perfectly heat conducting disk of radius $R$, which is uniformly heated on $\Delta T$, is in contact with a semi-infinite FE, which temperature is $T_0$ very far from the surface $z = 0$, and neglect the heat flux between the FE and the ambient air or vacuum. For the case the static temperature field $T(\vec{r}, t)$ and its vertical gradient inside the FE are given by expressions:

$$T(r,z) = T_0 + \frac{2}{\pi}\Delta T \arcsin\left(\frac{2R}{\sqrt{(r-R)^2+z^2}+\sqrt{(r+R)^2+z^2}}\right), \quad (0 \leq z \leq \infty) \quad \text{(A.1a)}$$

$$\frac{dT}{dz} = \frac{4Rz\Delta T}{\pi\sqrt{(r-R)^2+z^2}\sqrt{(r+R)^2+z^2}\sqrt{\left(\sqrt{(R-r)^2+z^2}+\sqrt{(R+r)^2+z^2}\right)^2 - 4R^2}}, \quad \text{(A.1b)}$$

where the polar radius $r = \sqrt{x^2 + y^2}$ is introduced. The integral form of Eq.(A.1a) is $T(r,z) = T_0 + \frac{2}{\pi}\Delta T \int_0^\infty e^{-kz} J_0(kr)\frac{\sin(kR)}{k} dk$. Note, that the characteristic time of temperature gradient relaxation is the minimum between the "vertical" time $h^2/\kappa_m^T$ and the "lateral" time $R^2/\kappa_m^T$.

**A1. Thermoelastic contribution to the PFM response.** This contribution comes from the inhomogeneous thermal expansion of the FE caused by the contact with a heated tip, and the expansion is proportional to the specific integral convolution of the material thermal expansion tensor with the temperature variation. We further restrict the analysis to the transversally isotropic thermal expansion tensor $\beta_{ij} = \delta_{ij}\beta_{ii}$ with $\beta_{11} = \beta_{22} \neq \beta_{33}$ ($\delta_{ij}$ is the Kroneker symbol). The maximal surface displacement corresponding to the point $z = 0$, i.e., the surface displacement at the tip-surface junction detected by SPM, is [98]:

$$u_3(x_1,x_2) = -\frac{1}{2\pi}\iiint_V \left(\begin{array}{c} \beta_{11}\xi_3 \frac{2(1+\nu)\left((x_1-\xi_1)^2+(x_2-\xi_2)^2\right)-(1-2\nu)\xi_3^2}{\left((x_1-\xi_1)^2+(x_2-\xi_2)^2+\xi_3^2\right)^{5/2}} \\ + \frac{3\xi_3^3 \beta_{33}}{\left((x_1-\xi_1)^2+(x_2-\xi_2)^2+\xi_3^2\right)^{5/2}} \end{array}\right) \vartheta(\vec{\xi}) \, d\xi_1 d\xi_2 d\xi_3 \quad \text{(A.2)}$$

where $\nu$ is the Poisson coefficient and $\vartheta(\vec{r}) = T(r,z) - T_0$. After Fourier transformation and using Percival theorem Eq.(A.2) becomes [98]:

$$u_3(r) = \int_0^\infty J_0(kr)k\,dk \int_0^\infty d\xi_3 e^{-k\xi_3}[\beta_{33}(1+k\xi_3) + \beta_{11}(1+2\nu-k\xi_3)]\vartheta(k,\xi_3). \quad \text{(A.3)}$$

Here $k = \sqrt{k_1^2 + k_2^2}$, $J_0(kr)$ is the Bessel function of zero order, and $\vartheta(k,\xi_3)$ is the 2D Fourier image of the temperature field $\vartheta(r,\xi_3)$ in the FE.

Using that $\vartheta(r,\xi_3) = \frac{2}{\pi}\Delta T \int_0^\infty e^{-k\xi_3} J_0(kr)\frac{sin(kR)}{k}dk$, it is easy to obtain the temperature variation Fourier image $\vartheta(k,\xi_3) = \frac{2}{\pi}\Delta T e^{-k\xi_3}\frac{sin(kR)}{k}$ that allows integration over $\xi_3$ and then over $k$ in Eq.(A.3). The answer is

$$u_3(r) = u_{TE}\int_0^\infty J_0(kr)\frac{sin(kR)}{k}dk = u_{TE}arcsin\left(\frac{2R}{|r-R|+|r+R|}\right), \quad (A.4)$$

where the factor $u_{TE} = \Delta T\frac{\beta_{11}(1+4\nu)+3\beta_{33}}{2\pi}$. As it was expected, the magnitude of the thermoelastic contribution to PFM response is proportional to the tip temperature variation $\Delta T$ and thermal expansion coefficients combination $\beta_{11}(1+4\nu)+3\beta_{33}$.

**A.2. Piezoelectric contribution to the PFM response.** In decoupled approximation, the surface displacement below the tip, $u_i(\mathbf{x})$, i.e., PFM signal in the point $\mathbf{x}$, can be rewritten as [100]:

$$u_i(\mathbf{x}) = \int_{-\infty}^\infty d\xi_1 \int_{-\infty}^\infty d\xi_2 \int_0^\infty d\xi_3 \frac{\partial G_{ij}(\xi_1,\xi_2,\xi_3)}{\partial \xi_k} e_{kjl}(x_1-\xi_1, x_2-\xi_2, \xi_3) E_l(-\xi_1,-\xi_2,\xi_3) \quad (A.5a)$$

Here $G_{ij}(\boldsymbol{\xi})$ is the Green's tensor, $e_{kjl}(\boldsymbol{\xi})$ are the strain piezoelectric tensor components representing material properties, which are proportional to the ferroelectric polarization $e_{kjl}(\boldsymbol{\xi}) \cong Q_{kjlm}P_m(\boldsymbol{\xi})$, and $E_l(\mathbf{x})$ is the electric field produced by the tip in the FE. Coordinate systems $\mathbf{x}$ and $\boldsymbol{\xi}$ are linked to the tip apex. For the most inorganic FE, the Green's tensor $G_{ij}(\mathbf{x}-\boldsymbol{\xi})$ can be approximated by the one of elastically isotropic half-plane, and is listed by Lur'e [112], and Landau and Lifshitz [113]. The in-plane polarization components are proportional to the acting electric field, and the out-of-plane component is given by Eqs.(3).

As a rough estimate for a vertical piezo-response of a uniaxial ferroelectric, we can use an expression:

$$u_3(\mathbf{x}) \cong Q \int_{-\infty}^\infty d\xi_1 \int_{-\infty}^\infty d\xi_2 \int_0^\infty d\xi_3 \frac{\partial G_{33}(\boldsymbol{\xi})}{\partial \xi_3} P_3(x_1-\xi_1, x_2-\xi_2, \xi_3) E_3(-\xi_1,-\xi_2,\xi_3), \quad (A.5b)$$

where $Q$ is a corresponding electrostriction coefficient; $G_{33}(\boldsymbol{\xi}) = \frac{1+\nu}{2\pi Y}\left(\frac{2(1-\nu)}{\xi} + \frac{\xi_3^2}{\xi^3}\right)$, where $\xi = \sqrt{\xi_1^2 + \xi_2^2 + \xi_3^2}$, $Y$ is Young's modulus, and $\nu$ is the Poisson ratio.

Using a disk-contact model, the electric potential $\varphi(r,z)$ and the field component $E_z(r,z)$ produced by the disk contact area are given by expressions

$$\varphi(r,z) \approx \frac{2}{\pi}U arcsin\left(\frac{2R}{\sqrt{(r-R)^2+(z/\gamma)^2}+\sqrt{(r+R)^2+(z/\gamma)^2}}\right), \quad (0 \leq z \leq \infty) \quad (A.6a)$$

$$E_z(r,z) = \frac{4R(z/\gamma)U}{\pi\sqrt{(r-R)^2+(z/\gamma)^2}\sqrt{(r+R)^2+(z/\gamma)^2}\left(\sqrt{(R-x)^2+(z/\gamma)^2}+\sqrt{(R+x)^2+(z/\gamma)^2}\right)^2-4R^2}, \quad \text{(A.6b)}$$

Here $U$ is the voltage applied to the tip, and $\gamma = \sqrt{\varepsilon_{33}/\varepsilon_{11}}$ is the dielectric anisotropy factor.

These assumptions allow us to regard thar all temperature-induced changes in polarization distribution are contained in the coefficient $\alpha_T(T(\vec{r}) - T_C)$ in Eq.(3a). It is known that, in the case of very smooth temperature gradient, the coefficient is responsible for the amplitude of the spontaneous polarization as $P_3(\vec{r}) \sim \sqrt{\alpha_T(T_C - T(\vec{r}))}$ and for the corresponding change of the domain wall "intrinsic" width $w(\vec{r}) \sim \frac{1}{\sqrt{\alpha_T(T_C - T(\vec{r}))}}$ in the ferroelectrics with the second order phase transition.

In particular, the vertical PFM response from the homogeneously polarized region is [84]

$$d_{33}^{eff} \approx \left(\frac{1}{4}+v\right)d_{31} + \frac{3}{4}d_{33} + \frac{d_{15}}{4}. \quad \text{(A.7a)}$$

The expression is formally the same as for a "cold" piezoresponse, but here the piezoelectric coefficients $d_{ij}$ have another sense. They are not bulk coefficients, but contain the information about the temperature variation $d_{ij} \sim \bar{P}_S$, where the averaging is performed over the region of tip filed concentration. Very approximately, the averaging is given by the integration, $\int_0^h dz \int_0^R r dr\, f$ is the case of very smooth temperature variation.

## Appendix B. The influence of the flexoelectric coupling

**Figures A1a-b** and **4d-e** are calculated without flexoelectric coupling, e.g., for $F_{ij} = 0$. **Figures A1c** and **A1f** are calculated for nonzero flexoelectric coefficients, $F_{ij} \neq 0$, which are listed in **Table I**. It is seen from the comparison of the figures, that the flexoelectric coupling does not influence on the displacement distribution, even tiny features are absent. This result may look unexpected, because the flexoelectric effect looks negligibly small even at the diffuse domain walls shown in **Fig. A1d**. However, it can be explained in the following way: for the heating on $\Delta T > 5$ K and nonzero voltages $|V| > 0.1$ V the electrostriction contribution strongly dominates over the flexoelectric contribution, and over the thermopolarization contribution too.

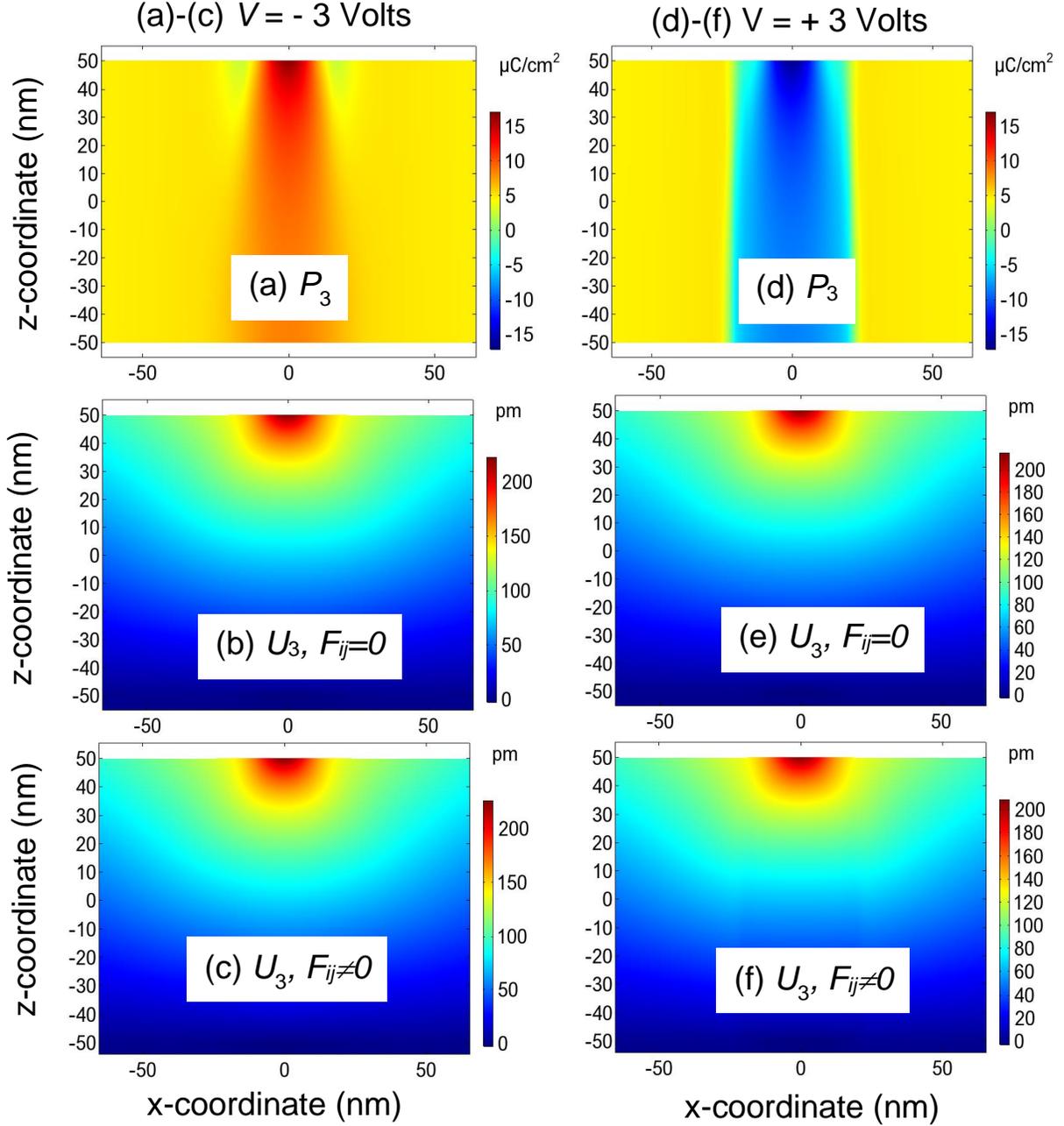

**FIG. A1**. Spatial distributions (xz cross-sections) of the polarization $P_3$ **(a, d)** and vertical displacement $U_3$ **(b, c, e, f)** of a thick SPS layer calculated for applied voltage $V = -3$ V **(a-c)** and $V = +3$ V **(d-f)**, tip-surface contact radius $R = 10$ nm, tip overheating $\Delta T = 50$ K, and $T_0 = 293$ K. Before heating the FE was homogeneously polarized. Plots **A1a-b** and **A1d-e** are calculated for $F_{ij} = 0$. Plots **A1c** and **A1f** are calculated for $F_{ij} \neq 0$ listed in **Table I**. Other parameters are also listed in **Table I.**